\documentstyle[aps,epsf,preprint]{revtex}
\tightenlines
\begin{document}

\renewcommand{\thesubsection}{\arabic{section}.\arabic{subsection}}
\renewcommand{\theequation}{\arabic{section}.\arabic{equation}}
\renewcommand{\thefootnote}{*}
\draft
\title{\Large {Alternative mass shell renormalization \\for minimal supersymmetric Higgs sector}}
\author{ Lian-You Shan\footnote{Email:~~~shanly@itp.ac.cn}}
\address{ CCAST (World Laboratory) P.O. Box 8730, Beijing 100080, China\\
Institute of Theoretical Physics, 
        Academia Sinica, \\P.O. Box 2735, Beijing 100080, China}


\date{*}
\maketitle  

\begin{abstract}
An Aoki-Denner form of the renormalization scheme is suggested for the physical amplitudes in MSSM. 
The Higgs sector is reparameterized with the mass of the CP odd scalar, and the mass of the heavy CP even one instead of $tan\beta$ in our scheme. 
The counterterms of $tan\beta$ is fixed perturbatively on mass shell just within Higgs sector. 
The counterterms of gauge-scalar mixings are defined with Ward-Takahashi identities from scalar-scalar mixings.
The effect of the reparameterization is also probed on the radiative correction of the mass of the lightest Higgs.
\end{abstract}

\section{Introduction}
\vspace*{0.1cm} \hspace*{0.5cm}
\renewcommand{\thepage}{\arabic{page}}

~~~ 
In the minimal supersymmetric standard model (MSSM)\cite{mssm}, the masses and 
couplings of physical bosons are restricted and can be expressed 
in terms of merely two free parameters at tree level, although supersymmetry is softly broken\cite{guide}. 
Those constrains no doubt will be affected by radiative corrections, especially by 
top/stop quark loops. 

If MSSM is a perturbative theory as reckonable as it had been expected, its qualities at lower 
order should be kept somehow up to higher order. 
For example, if those treelevel relations had been disregarded completely, the prediction for the mass of the light CP even Higgs $M_h<130~GeV $ in the effective potential(EP) approach 
\cite{ep} or in the renormalization group method(RG) \cite{rg} will be non-trivially complicated. 
These kinds of work gave a logarithm correction 
$\varepsilon \sim 3G_F M^4_t \sin^2\beta \log(1+m^2_{\tilde t} / m^2_t)$\cite{ep}  
to estimate the mass of Higgs bosons with good approximation, which can be used for Higgs phenomena\cite{spira}.

When one counts in the momentum dependence of the full set of Green functions, one can also deduce a logarithm 
correction like that $\varepsilon$ within the framework of Feynman diagrammatic calculation (FDC). 
At the same time, FDC is also well necessary for the phenomena on present or future 
colliders \cite{colli}, whose main goal is searching for the (lightest) Higgs boson. 
To identify a Higgs boson on wherever Tevatron, LHC or NLC, the knowledge of its mass is necessary indeed. 
Especially, to study whether the produced Higgs is a supersymmetric one, an appropriate 
supersymmetry-like simulation for its production cross sections and decay widths is more crucial and beneficial.

Then, instead of abandoning all the tree level relations in MSSM, 
one should investigate which of those simple 
supersymmetric constrains can be remained perturbatively in FDC and how the other variables can be deduced 
loop by loop.
With this bias we noticed \cite{phC,aDa}, which have developed renormalization
procedures within the on-mass-shell scheme following \cite{hollik}. In their representation, 
the physical mass of Higgs boson was acquainted as the pole solution for the renormalized propagator, and the Higgs phenomena can be predicted systematically\cite{vdre}.
Recently, even the two loop FDC in the same line of \cite{hollik} had been 
developed for the prediction of the parameter $\rho=M^2_W / (M^2_Z \cos^2 \theta_w)$ 
and the mass of Higgs bosons \cite{SH}.  
All these works have demonstrated the efficacy of FDC. 

So in this work we try to spread the formalism in \cite{phC,aDa}, and highlight their effect as FDC for decay width or cross section.
We seek for an alternative realization motivated by \cite{aoki} and \cite{denner}, in which the wave function renormalization of mass eigenstates are performed explicitly, and in which the gauge fixing terms are renormalized simply, to offer a practical option for general MSSM perturbative calculations.
Such kind of frame has been established for the general two Higgs doublet model (2HDM), for example in \cite{santos}. However, the property of the supersymmetry in MSSM allows us to give more relations in the radiative corrections.
Similar consideration had ever been adopted in \cite{pierce} for radiative corrections.
 
In both \cite{pierce} and \cite{phC}, the $tan\beta$ was selected as an input parameter of MSSM and its counterterm is subtracted with a $\overline {MS}$ manner. Since $\delta \beta$ is much used for most FDC of MSSM, it should be fixed at a definite scale with an Ultraviolet (UV) finite part.
For the simplicity of loop calculations, it should also be defined through a set of 2-point one particle irreducible (1PI) Green functions but not the complicated 3-points Green functions. 
(It's well known that, in SM the gauge symmetry has simplified the counterterm of electric charge $\delta e$ as a combination of the self energies of neutral gauge bosons). 
Then the reparameterization in our scheme is a reasonable attempt through replacing $tan\beta$ with the heavy CP even scalar $M_H$.

In addition, the mixing of gauge and Higgs bosons raised outstanding since MSSM is a gauge theory with two scalar doublets as a special 2HDM. 
Especially the heavy top (super) quark will contribute a large correction to these mixing loops, which will be necessary 
for the physical process involving the pseudo-scalar or the charged Higgs bosons.
To our knowledge, this subject is less discussed as a part of systematic MSSM renormalization from the point view of gauge invariance, although various of treatments were already defined from the 
subtraction of Goldstone propagators in time of need. Since the Ward-Takahashi identities (WTI) plays an important role for the renormalization of 
gauge field theory, we tried to generalize the treatment in \cite{aoki} to this MSSM case for the counterterms of gauge-scalar mixings.

The present paper is organized as following. In section II, we introduced the conventions and notations for MSSM. 
In section III, we accomplished the Aoki-Denner form of the renormalization of MSSM, including the pole mass of the lightest CP even Higgs boson and the on-mass-shell counterterms of $\beta$. 
In section IV we deduced the wave-function renormalization constants of gauge-scalar mixing terms. 
A brief discussion is oriented on the 
application of these formulae in the last section. Some essential expressions are listed in the Appendices.

\section{ Tree level structure of MSSM and notations}
\setcounter{equation}{0}\setcounter{footnote}{0}

The original ${SU(2)}_L\otimes {U(1)}_Y$ gauge invariant Higgs sector in MSSM read, 
\begin{equation}
L_{kin}=
\overline H_1 D_{-~\mu}^\dagger D^\mu_- H_1 + \overline H_2 D_{+~\mu}^\dagger D^\mu_+ H_2
\end{equation}
where $D^\mu_\mp=\partial^\mu \mp {i\over 2}g_1 B^\mu -ig_2T^a W^{a~\mu}$.~~
For low energy phenomena, the Higgs sector of MSSM has a soft broken potential with explicit $CP$ conservation,
\begin{eqnarray}
V_{soft} &=& m_1^2 \overline H_1 H_1+ m_2^2 \overline H_2 H_2 - m_{3}^2 ( \epsilon_{ab} H_1^a H_2^b + h.c. ) \nonumber \\
&+& \frac{1}{8} g^2 ( \overline H_1 H_1- \overline H_2 H_2 )^2 - \frac{g_2^2}{2} | \overline H_1 H_2 |^2 
\label{soft}
\end{eqnarray}
where $m^2_{3}$ is defined to be negative and $\epsilon_{12}=-\epsilon_{21}=-1,~g^2=g^2_1+g^2_2$.
Here can we count clearly the five parameters of this model,
$g_1, g_2, m_1, m_2$, and $m_3$ (where a $\mu$ has been absorbed into $m_1$ and $m_2$).
This model in Higgs sector has fewer parameters than 2HDM, so it should be more predictive.

Down to Electroweak scale, Higgs fields develop their vacuum expectation value (VEV) $v_1\neq 0,v_2\neq 0$ 
 and mix into mass eigenstates. With the same components of $\Phi_1, \Phi_2$ in \cite{kane}, 
the doublets are,
\begin{eqnarray}
H_1 = \left( \begin{array}{c} H_1^1 \\ H_1^2 \end{array} \right) =
      \left( \begin{array}{c} (v_1 + \phi_1^{0} - i \chi_1^{0})/\sqrt{2} \\
 - \phi_1^-       \end{array}  \right)  \ , ~~
H_2 = \left( \begin{array}{c} H_2^1 \\ H_2^2 \end{array} \right) =
      \left( \begin{array}{c} \phi_2^+ \\
      (v_2 + \phi_2^0 + i \chi_2^0)/\sqrt{2} \end{array} \right) \  \\
\end{eqnarray}
but neither $v_1$ nor $v_2$ is new independent parameters. 
They can be induced as functions of the five original parameters from the minimum point of the potential\cite{haber}
\begin{equation}
\frac {\partial V}{\partial v_1}=0,~~~~\frac {\partial V}{\partial v_2}=0,~~(all~fields \rightarrow 0) 
\label{minim}
\end{equation}
and $v\equiv \sqrt{v_1^2+v_2^2}$ gives the mass of gauge bosons as known well in SM.

Furthermore we choose the mass of pseudo-scalar Higgs $M_A$ and the mass of heavier CP even neutral Higgs $M_H$ as the input parameters
for the Higgs sector. Then other parameters can be represented upon these five independent
parameters.
\begin{equation}
\cos^2 2\beta=\frac{M^2_H (M^2_A + M^2_Z -M^2_H)}{M^2_A M^2_Z}
\label{betadef}
\end{equation}
\begin{eqnarray}
& &\tan 2\alpha=\tan 2\beta \frac{M_A^2+M_Z^2}{M_A^2-M_Z^2},~~
m^2_3=-M_A^2 \sin \beta \cos \beta, \nonumber \\
& &v_1=v \cos \beta,~~~v_2=v \sin \beta
\label{v1v2v3}
\end{eqnarray}
We'll see later that Eq.(\ref{betadef}) and the first of Eq.(\ref{v1v2v3}) can be used to define $\beta$ and $\alpha$ loop by loop.
At tree-level, Eq.~(\ref{minim}) appear explicitly
\begin{eqnarray}
&& 0= m^2_3 v \sin\beta + v \cos \beta (m^2_1 + M_Z^2 \cos 2\beta /2)\nonumber \\ && 
0= m^2_3 v \cos \beta + v \sin \beta (m^2_2 - M_Z^2 \cos 2\beta /2) 
\label{tremin}
\end{eqnarray}
which has used Eq.~(\ref{betadef}, \ref{v1v2v3}), and implied that, $m^2_1, m^2_2$ can be 
considered as functions dependent on $M_A, M_H, M_Z, M_W$ and $e$.
This constrain will be changed by loop correction although they lead to simple 
tree-level mass for Higgs bosons,
\begin{equation}
M^2_h = \frac{1}{2} [~~M_A^2 + M_Z^2 - \Delta ],~~~M_{H^+}^2 = M_A^2 + M_W^2 
\label{treemass}
\end{equation}
where 
$\Delta=\sqrt{ ( M_A^2 +M_Z^2 )^2 - 4 M_Z^2 M_A^2 \cos ^2 2\beta }$ 

The gauge-Goldstone mixing terms must be encountered when the gauge invariant eigenstates are transformed into mass eigenstates. 
For example, to $Z$ boson there is 
one term $  Z^{\mu} \partial_{\mu} G $ with a coefficient 
$(v_1 \cos \beta + v_2 \sin \beta) \sqrt{g_1^2+g_2^2}/2$. In our selecting of parameterization, 
this mixing term becomes
\begin{equation}
{\cal L}_{mix}=-M_Z Z^{\mu} \partial_{\mu} G  
\label{zgmix}
\end{equation}
Fortunately, this mixing can be cancelled at tree level by the one from so-called gauge-fixing 
term,
\begin{equation}
{\cal L}_{gf}=-\frac{1}{2\alpha_z}{(\partial_{\mu} Z^{\mu}+{\alpha}_z M_Z G)}^2
\label{zgf}
\end{equation}
which is necessary for the quantization to gauge fields. The same thing keep for 
W gauge boson and photon. 

Here we have chosen the SM-like gauge fixing,
and in following calculation, we adopt the 't Hooft-Feynman 
gauge, ${\alpha}_z={\alpha}_w={\alpha}_\gamma=1$, since the physical result should be gauge independent. 

\section{renormalization procedure for MSSM}
\setcounter{equation}{0}\setcounter{footnote}{0}

A procedure of renormalization is expected to perform the perturbative calculation. 
One choice is naively including the virtual super particles into the radiative loops in \cite{santos} 
and mechanically applying the subtraction formulae listed there, with an argument that the tree-level relations 
are spoilt. 
Unless the necessity to withdraw so far, however, we prefer to find one mediocre formalism aimed at Eq.~(\ref{soft}). The supersymmetric structure of this potential still enable us to predict perturbatively the pole mass of the lightest Higgs and to define the counterterm of $\beta$ even though the conventions of \cite{aoki,denner} are followed here.
\subsection{general framework}
This scheme defines explicitly the renormalization constants of fields as mass eigenstates,
\begin{eqnarray}
& & W^{\pm}_{\mu}\rightarrow Z^{1/2}_W W^{\pm}_{\mu}~,~~
  \left(\matrix{Z_{\mu}\cr A_{\mu}}\right)
  \rightarrow \left(\matrix{ Z^{1/2}_Z & Z^{1/2}_{Z\gamma}
                        \cr     Z^{1/2}_{\gamma Z} & Z^{1/2}_{\gamma }}\right)
  \left(\matrix{Z_{\mu}\cr A_{\mu}}\right) \nonumber \\
& & \left(\matrix{ A \cr G }\right)
  \rightarrow \left(\matrix{ Z^{1/2}_A & Z^{1/2}_{AG}
                       \cr     Z^{1/2}_{GA} & Z^{1/2}_G }\right)
  \left(\matrix{A \cr G}\right) ,~~  \left(\matrix{ H \cr h }\right)
  \rightarrow \left(\matrix{ Z^{1/2}_H & Z^{1/2}_{Hh}
                       \cr      Z^{1/2}_{hH} & Z^{1/2}_h }\right)
  \left(\matrix{H \cr h}\right)
\label{wave}
\end{eqnarray}

We haven't taken the renormalization to the gauge eigenstates used in \cite{phC} and
\cite{aDa} such as,
\begin{eqnarray}
& &  H_i\rightarrow Z^{1/2}_{Hi} H_i,~~~ B_{\mu}\rightarrow Z^{1/2}_B B_{\mu},~~~
\stackrel{\rightarrow}{W_{\mu}}\rightarrow Z^{1/2}_W \stackrel{\rightarrow}{W_{\mu}} \nonumber\\
& &
  (\xi^{B,W}_{1,2} \rightarrow 1+\delta \xi^{B,W}_{1,2})
\label{gaupara}
\end{eqnarray} 
which seems more compact and concise, so we have to seek alternative way to treat tadpoles
and define $\delta\beta$.  As to the input parameters, our scheme prefer the renormalization 
to the five physical parameters bellow,
\begin{eqnarray}
 M_Z^2 & \rightarrow & M_Z^2 + \delta M_Z^2 ~~~
 M_W^2 \rightarrow M_W^2 + \delta M_W^2  \nonumber \\
 M_A^2 & \rightarrow & M_A^2 + \delta m_A^2 ~~~
 M_H^2  \rightarrow  M_H^2 + \delta m_H^2 \nonumber \\ 
 e     & \rightarrow & e + \delta e  
\label{renpara}
\end{eqnarray}
where $M_W,~M_Z$ are the mass of gauge bosons, $e $ is the electric charge, and they are renormalized in
the conventional electroweak treatment
\begin{eqnarray}
& &\Re e\ \hat{\Sigma}_Z (k^2) =  {\Sigma}_Z (k^2) - \delta M_Z^2 + \delta Z_Z
( k^2 - M_Z^2 ) =0 \nonumber \\
& & \Re e\ \hat{\Sigma}_W (k^2)  =  {\Sigma}_W (k^2) - \delta M_W^2 + \delta Z_W
( k^2 - M_W^2) =0  \nonumber \\
& & \hat{\Gamma}_\mu^{\gamma ee} ( k^2=0,\not{p}=\not{q}=m_e)  =  i e\gamma_\mu
\label{gaushell}
\end{eqnarray}

Regarded as the physical mass of the pseudo-scalar (heavy CP even neutral) Higgs boson, the $M_A$ ($M_H$) in 
Eq.~(\ref{renpara}) finds its counterterm in a way like Eq.~(\ref{gaushell}), as will be demonstrated later. 

In addition to the wave function renormalization of charged Higgs, the renormalization of 
physical Higgs masses,
\begin{equation}
M_h^2  \rightarrow  M_h^2 + \delta M_h^2 ~~~~
M^2_{H^+} \rightarrow M^2_{H^+}  + \delta  M^2_{H^+}
\label{higmas}
\end{equation}
formally complete the renormalization of 2HDM including MSSM. In the conventional treatment \cite{sola} and
 \cite{santos}, Eq.~(\ref{higmas}) fix the counterterms of Higgs mass, and gives no information on the value 
of these masses. However, in MSSM we can give alternative interpretation to Eq.~(\ref{higmas}) 
and subtract the magnitude of Higgs masses from it.

\subsection{constrains on mass counterterms}
In MSSM, the mass relations between gauge and Higgs bosons are connected by Eqs.~(\ref{tremin}) and Eq.~(\ref{betadef},\ref{v1v2v3}). Since the renormalization should not increase the number of the independent (free) parameters, we have to reproduce the masses of bosons through the breaking of gauge symmetry, to investigate how those connections can be regulated by the loop corrections. In another word, the relation between different counterterms should be determined by these constrains. Then the rescaling of the scalar's VEV ($i.e.$ the renormalization of those constrains) is of the first importance. For the convenience to expound, we construct the generalized form of Eqs.~(\ref{tremin})
\begin{eqnarray}
{\cal T}_h=\frac{v}{8}
\{[8 m^2_3 \cos\beta + \sin\beta (8 m^2_2 - g^2 v^2 \cos 2\beta )] \cos \alpha \nonumber\\
- [8 m^2_3 \sin\beta +\cos\beta (8 m^2_1  + g^2 v^2 \cos 2\beta) ]\sin \alpha \} \nonumber\\
{\cal T}_H=\frac{v}{8}
\{[8 m^2_3 \sin\beta +\cos\beta  (8 m^2_1  + g^2 v^2 \cos 2\beta)] \cos \alpha \nonumber\\
+ [8 m^2_3 \cos\beta + \sin\beta (8 m^2_2 - g^2 v^2 \cos 2\beta )] \sin \alpha\}
\label{newmin}
\end{eqnarray}     
The $v_1, v_2$ should generate properly the masses of gauge bosons and fall into the last two of 
Eq.~(\ref{v1v2v3}) order by order. To generate the masses of Higgs bosons, it's convenient to employ the original matrices 
in the quadratic terms of scalar fields. The matrices (a factor $\frac{1}{2}$ compressed) have been rotated but are not necessarily identified 
as the physical masses. The most important one is the mass form of the pseudo-scalar $A$,
\begin{equation}
{\cal M}_{AA}=2[~(-4m^2_1 + 4m^2_2-g^2 v^2 \cos 2\beta )\cos 2\vartheta +4(m^2_1 +m^2_2 -2m^2_3 \sin 2\vartheta)~]/16
\label{Amass}
\end{equation} 
which has no more impact but a equation to solve $m^2_1, m^2_2, m^2_3$, associated with Eq.~(\ref{newmin}).
It's easy to check from the Eq.~(\ref{ms}) in the Appendices that, $m^2_1, m^2_2, m^2_3$ return to their tree-level form Eqs.~(\ref{tremin}) and 
~(\ref{betadef}, \ref{v1v2v3}) only if $\vartheta \rightarrow \beta, {\cal T}_h \rightarrow 0,{\cal T}_H \rightarrow 0$. 
Then the mass matrices (quadratic form) of the Higgs sector can be recast as functions of $e, M_W, M_Z, M_A, \beta(M^2_H)$. 
  
These fundamental parameters ($e~,M_W, M_Z, M_A, M_H$) can run from their bare to corresponding renormalized (physical) values, as described 
in Eq.~(\ref{renpara}). When such a replacement is performed, a natural renormalization condition show itself up,
\begin{eqnarray}
& &{\cal T}^R_h\equiv {\cal T}_h(e_R, {M_W}_R, {M_Z}_R, {M_A}_R, {M_H}_R) =0\nonumber \\
& &{\cal T}^R_H\equiv {\cal T}_H(e_R, {M_W}_R, {M_Z}_R, {M_A}_R, {M_H}_R)=0
\label{tadren}
\end{eqnarray}
This indicates nothing else but that, the physical rescaling of VEV is to eliminate the linear 
terms of Higgs fields, so that each renormalized (one point) Green function has tree level form in renormalized parameters, and 
so that $v_{iR}$ is the place where the potential reach its minimum, and the resonable relation 
\begin{equation}
v_1\delta v_1+v_2\delta v_2=v\delta v,~~~ \frac{v_2}{v_1} (\frac{\delta v_2}{v_2}-\frac{\delta v_1}{v_1})=\sec^2 \beta~\delta\beta
\label{vcoun}
\end{equation}
rather than $\delta v_2=\delta v_1=\delta v=0$ had been used. From now on, the subscription R is omitted 
on the right side of equations unless the renormalized quantity is not equivalent to the physical one. 
To show how $\delta\beta$ is traded for $\delta M^2_H$, we write the other two transformations at length
\begin{eqnarray}
& &{\cal M}_{AA}(e, M_W, M_Z, M_A, M_H)\rightarrow M^2_A +\delta M^2_A \nonumber\\
& &{\cal M}_{HH}(e, M_W, M_Z, M_A, M_H) \equiv [(4 m^2_1 + 4 m^2_2 + g^2 v^2) \nonumber \\
& & +2 (2 m^2_1 - 2 m^2_2 +  g^2 v^2 \cos 2\beta ) \cos 2\alpha + 2 (4 m^2_3 - g^2 \sin\beta \cos\beta ) \sin 2\alpha ]/8\nonumber\\
& &= \frac{1}{2} [ M_A^2 + M_Z^2 +\Delta ]+{\cal T}_{HH} \rightarrow {\cal M}^R_{HH}+\delta {\cal M}_{HH} \nonumber\\
& &=[~M^2_H + {\cal T}^R_{HH}~]+\{~[~(\frac{1}{2}+\frac{\partial\Delta}{\partial M^2_A}) \delta M^2_A
+(\frac{1}{2}+\frac{\partial\Delta}{\partial M^2_Z}) \delta M^2_Z
+\frac{\partial\Delta}{\partial\beta}\delta\beta~] + T_{HH}~\}\nonumber\\
& &=[~M^2_H +{\cal T}^R_{HH}~]+\{~[~\delta M^2_H~] + T_{HH}\}
\label{Hpolemass}
\end{eqnarray}
where ${\cal T}_{HH}$ is a linear combination of ${\cal T}_h$,~${\cal T}_H$ and ${\cal T}^R_{HH}=0$. 
Then Eqs.~(\ref{vcoun},~\ref{Amass}) ~and 
\begin{eqnarray}
{\cal T}_h(e, M_W, M_Z, M_A, M_H)\rightarrow {\cal T}^R_h + T_h=0+ T_h, \nonumber\\
{\cal T}_H(e, M_W, M_Z, M_A, M_H)\rightarrow {\cal T}^R_H + T_H=0+ T_H
\end{eqnarray}
induced
\begin{eqnarray}
& &{\cal M}_{AG}(e, M_W, M_Z, M_A, M_H)\rightarrow [~{\cal M}^R_{AG}~] + \delta {\cal M}_{AG} \nonumber\\
& &=[~\frac{M^2_A (v_2 \cos\vartheta - v_1 \sin\vartheta)}{v_1 \cos\vartheta +v_2 \sin\vartheta}+{\cal T}^R_{AG}~]+ T_{AG}\nonumber\\
& &{\cal M}_{hh}(e, M_W, M_Z, M_A, M_H)\rightarrow M^2_{hR} + {\cal T}^R_{hh}+ [~\delta M^2_h +T_{hh}~]
\label{AAGcoun}
\end{eqnarray}
where $M^2_{hR}$ obey Eq.~(\ref{treemass}) and $\delta {\cal M}_{hh}$ is just its variation,
\begin{equation}
\delta {\cal M}_{hh}\equiv \delta M^2_h + T_{hh}= T_{hh}+(\frac{1}{2}-\frac{\partial\Delta}{\partial M^2_A}) \delta M^2_A+(\frac{1}{2}-\frac{\partial\Delta}{\partial M^2_Z}) \delta M^2_Z
-\frac{\partial\Delta}{\partial\beta}\delta\beta 
\label{dh}
\end{equation}

Here the rotational matrix had been defined with different angles as shown in the Eq.~(\ref{angles}).
An easy algebra concludes that, $\delta\vartheta$ (or $\delta\vartheta_+$) can be cancelled automatically and neatly by $\delta\beta$, only if $\beta =\vartheta = {\vartheta}_+$ is set in the coefficients of these counterterms. 
This pleasing result indicates that, in this scheme, the angles for Higgs 
coupling to other particles can be kept as only one angle, $i.e.$ $\beta$. And, one $\delta \beta$ is sufficient and consistent for any one loop calculations. 

\subsection{$M^2_{hP}$ as pole mass }

Combined with the fields renormalization in Eq.~(\ref{wave}), above mass counterterms can be fixed by the so-called on mass shell renormalization conditions in the Higgs sector,

1). Tadpoles
\begin{eqnarray}
0={\cal T}^R_h +  T_h + t^h ,~~~
0={\cal T}^R_H + T_H + t^H 
\end{eqnarray}
2). Heavy neutral $CP$ even Higgs
\begin{eqnarray} 
& & \frac{d }{dq^2} \Sigma_{HH} (q^2) |_{q^2=M^2_H}~+ Z_{HH} + Z_{Hh} = 1\nonumber\\
& & \Re e\ \hat{\Sigma}_H (q^2)|_{q^2=M^2_H}=[~\Sigma_{HH}(q^2) +(Z_H +Z_{hH}) q^2~]|_{q^2=M^2_H}-2 Z^{1/2}_H Z^{1/2}_{Hh}\delta{\cal M}_{Hh} \nonumber\\
& & -Z_H ({\cal M}_{HH}+\delta {\cal M}_{HH})- Z_{hH}({\cal M}_{hh} +\delta {\cal M}_{hh})=0\nonumber\\
& & \Re e\ \hat{\Sigma}_{Hh} (q^2)|_{q^2=M^2_H}=[~\Sigma_{Hh}(q^2)+(Z^{1/2}_{Hh} Z^{1/2}_H + Z^{1/2}_h Z^{1/2}_{hH}) q^2~]|_{q^2=M^2_H}- Z_{hH}^{1/2} Z^{1/2}_h ({\cal M}_{hh} +\delta {\cal M}_{hh})\nonumber\\
& &- Z^{1/2}_{Hh} Z^{1/2}_H ({\cal M}_{HH}+\delta {\cal M}_{HH})-(Z^{1/2}_h Z^{1/2}_H+Z^{1/2}_{Hh}Z^{1/2}_{hH})\delta{\cal M}_{Hh}=0
\label{heavy}
\end{eqnarray}
3). light neutral $CP$ even Higgs
\begin{eqnarray}
& & \frac{d }{dq^2}  ~\Sigma_{hh} ( q^2)+ Z_{hh} + Z_{hH} = 1\nonumber\\
& & \Re e\ \hat{\Sigma}_{hh} (q^2)=[~\Sigma_{h}(q^2) +(Z_h +Z_{Hh}) q^2 ~]-Z_h ({\cal M}_{hh} +\delta {\cal M}_{hh})\nonumber\\
& &- Z_{Hh} ({\cal M}_{HH}+\delta {\cal M}_{HH})-2 Z^{1/2}_h Z^{1/2}_{Hh}\delta{\cal M}_{Hh}=0\nonumber\\
& & \Re e\ \hat{\Sigma}_{hH} (q^2)=[~\Sigma_{Hh}(q^2)+(Z^{1/2}_h Z^{1/2}_{hH}+Z^{1/2}_{Hh} Z^{1/2}_H) q^2 ~]-Z_{hH}^{1/2} Z^{1/2}_h ({\cal M}_{hh} +\delta {\cal M}_{hh})\nonumber\\
& &-Z^{1/2}_{Hh} Z^{1/2}_H ({\cal M}_{HH}+\delta {\cal M}_{HH})-(Z^{1/2}_H Z^{1/2}_h+Z^{1/2}_{Hh}Z^{1/2}_{hH})\delta{\cal M}_{Hh}=0
\label{light}
\end{eqnarray}
The subtraction for the neutral $CP$ odd Higgs $A$ has the same form as Eq.~(\ref{heavy}) when a particle 
substitution $(H,h)\rightarrow (G,A)$ is made. The symbolics $t^h, t^H, \Sigma_{HH}, \Sigma_{hh},\Sigma_{hH}, \Sigma_{AA}$ and $\Sigma_{GA}$ denote 
the 1PI Green functions ( loop integrals with UV divergence ). For the sector that concerns the selected input parameters it's easy to solve,
\begin{eqnarray}
& &T_h = - t^h ,~~ T_H = - t^H \nonumber\\
& &\delta M_H^2= \Sigma_{HH} ( M^2_H)-T_{HH},~~\delta Z_H=-\frac{d}{dq^2} \Sigma_{HH} ( M^2_H)\nonumber\\
& &\delta Z_{hH}=\frac{2 [~T_{hH}-\Sigma_{hH} ( M^2_H)~]}{M^2_H-M^2_{hR}}\nonumber\\
& &\delta M_{A}^2= \Sigma_{AA} ( M^2_A), ~~\delta Z_A=-\frac{d}{dq^2} \Sigma_{AA} ( M^2_A),~~\delta Z_G=-\frac{d}{dq^2} \Sigma_{GG}(0) \nonumber\\
& &\delta Z_{GA}=\frac{2 [~T_{GA}-\Sigma_{GA} ( M^2_A)~]}{M^2_A} 
\label{couns}
\end{eqnarray}
where Eqs.~(\ref{Hpolemass}) and ~(\ref{AAGcoun}) have been used.

It's noticeable that, the variable of $\Sigma_{HH}$ is the physical mass $M^2_H$ in Eq.~(\ref{heavy}), which is just $M^2_{HR}$ at the same time. 
On the contrary the variable of $\Sigma_{hh}$ is not $M^2_{hR}$ in Eq.~(\ref{light}). 
The physical (pole) mass of the light Higgs can be solved as function of $M^2_A, M^2_H, 
M^2_Z, e$. One Taylor expansion may simplify the analysis and help us to realize this point.
\begin{eqnarray} 
& &q^2=M^2_{hR}+\frac{1}{2}{(q^2-M^2_{hR})}^2\frac{d^2}{d^2 q^2}\Sigma_{hh} (q^2)|_{q^2=M^2_{hR}}+{\cal O}({(q^2-M^2_{hR})}^3)\nonumber\\
& &+\frac{1}{2}[~\delta M^2_Z +\delta M^2_A-\delta \Delta~]-\Sigma_{hh} (M^2_{hR})+T_{hh}
\label{hbeta}
\end{eqnarray} 
The choice $q^2=M^2_{HR}$ in Eq.~(\ref{heavy}) makes Eq.~(\ref{light}) independent of $\delta \Delta$ ($i.e.~\delta \beta$) when 
these two equations are added together,
\begin{eqnarray}
q^2&=&M^2_Z+M^2_A-M^2_H+\frac{1}{2}{(q^2-M^2_{hR})}^2\frac{d^2}{d^2 q^2}\Sigma_{hh} (q^2)|_{q^2=M^2_{hR}}+{\cal O}({(q^2-M^2_{hR})}^3)\nonumber\\
& &-[~\Sigma_{hh} (M^2_{hR})+\Sigma_{HH} (M^2_H)~]+[~\delta M^2_Z +\delta M^2_A]+[~T_{hh}+T_{HH}~]
\label{hpolemass}
\end{eqnarray} 
The $\frac{d^2}{d^2 q^2}\Sigma_{hh} (q^2)|_{q^2=M^2_{hR}}$ term is UV finite unless the order of 
divergence in the self energies of the scalars were higher than quadratic.
Had not the supersymmetry been broken, the remained $\Sigma_{hh} (M^2_h), \Sigma_{HH}(M^2_H), 
\Sigma_{AA}(M^2_A), \Sigma^T_{ZZ}(M^2_Z)$ and $ T_{HH}, T_{hh}$ would also be convergent in a super-renormalizable 
theory. Although the breaking of supersymmetry cause those self-energies divergent, we can put forward a question, whether 
there remains a space to accommodate the cancelation of all of those divergence.

Fortunately, the possibility for the last line in Eq.~(\ref{hpolemass}) to be UV finite, had been hinted in the Append.E.7 of \cite{mssm} and verified analytically in \cite{gunion}.   
Our combination of self-energies for $Z^\mu$, $A,~H$ and $h$ bosons in Eq.~(\ref{hpolemass}) had been employed as the ``renormalization of the neutral Higgs boson mass sum rule" of \cite{pierce,berger}. 

We have examen it with the top quark and its squarks in one loop corrections. We can manifest the UV divergence in these $ 4\times 7 + 2\times 3 = 36 $ diagrams to cancel neatly.

\subsection{the FREE $M_H$ and the counterterm of $\beta$}
A cautious one will notice the using of in Eq.~(\ref{betadef}) for Eq.~(\ref{hpolemass}). This means that we 
have defined $\beta$ as an induced variable perturbatively through Eq.~(\ref{betadef}), $i.e$
\begin{equation}
 M^2_H - M^2_Z < M^2_A < M^2_H  
\label{sequen}
\end{equation}

Such a presumption originates mainly from that, no experiment has indicated the necessary to abort this relation till 
now, although LEP does not prefer a SM Higgs lighter than gauge boson $Z$ \cite{lep}. The tendency $ \lim_{M_A \rightarrow \infty} M_H = M_A  $ is one prediction of MSSM at tree level and it is found to hold in the evolutions at one loop
\cite{guide, fdc}. Here we consider it as a possible point of MSSM to test.

There is no theoretical evaluation against (\ref{sequen}) yet. In reality, we could have erased the subscription $q^2=M^2_H$ from Eq.~(\ref{heavy}) and could have got an expression analogous with Eq.~(\ref{hbeta}) for the 
heavy Higgs. Nevertheless, such two equations could never be sufficient for three variables ($M^2_{HP}, M^2_{hP}, \delta \Delta$) if we had employed $\beta$ as an independent input, although such a treatment sounds more strict and careful. 
Furthermore, the UV convergent part of $\delta \beta$ is difficult to tag although its UV divergent part (${\overline {MS}}$) had been fixed uniquely by the gauge symmetry, so that the uncertainty in $\delta \beta$ would be traded into $M^2_{hP},~M^2_{HP}$ by $\delta\Delta$ if $q^2=M^2_H$ was dismissed.
 In fact the $M^2_{HP}$ predicted in those $\beta -scheme $ is never destined to be conflict with Expression~(\ref{sequen}).

 The EP \cite{ep} performed little numerical evaluations for $M^2_{HP}$. One typical numerical evolution in RG approach can be found in the first of \cite{rg}, and expression~(\ref{sequen}) overlaps most of the permitted region in its "$Fig. 4-b, ~Bounds~on~the~higgs~masses$" when "$R=v_2/v_1 > 1 $".
 Even the most recent works with stop mixing and RG-improvement, \cite{effpoten}, \cite{apprada}, can not defeat the spectrum (\ref{sequen}) definitely. 
In that kind of language \cite{epslon}, our spectrum means $M^2_{hP}-M^2_Z < \varepsilon < M^2_{hP}$, which is a natural space.
The former FDC evaluations, the figures in \cite{fdc} confirmed the same spectra when $\tan\beta >1$. 

 Those arguments for the tiny of the UV convergent part in $\delta \beta$ is just the one for us to neglect the loop effect for the difference (violation) from Eq.(\ref{betadef}), although which follows the tree level assumption $tan\beta > 1 $. 

So we straightforward terminate using $\beta$ but start utilizing $M^2_H$ as a free input parameter by $M^2_H\equiv M^2_{HPhys}=M^2_{HR}$. 
The measurement of $M^2_H$ may be not as early as we expect, but its physical definition 
is always more unequivocal than the physical definition of $\beta$ itself. 
 Another reparameterization attempt had been made in \cite{phC}. There $\beta$ 
was replaced by the mass of the lightest Higgs boson $M^2_h$ which may be measured first. 

The counterterm of $M^2_H$ then brought us the one of $\beta$ through Eq.(\ref{betadef}),
\begin{equation}
\delta \beta=\epsilon \frac{\delta Z~ (H-A) H A}{M_{ZHA}}+\frac{\delta H~ (A+Z-2 H)A Z+\delta A~ (H-Z) H Z }{M_{ZHA}}
\label{db}
\end{equation}
where $\epsilon\equiv Sign(v_2-v_1)$, and 
\begin{equation}
\ M_{ZHA}={4 A Z \sqrt{H (H-Z)(H-A)(A+Z-H)}} \nonumber
\end{equation}
$A, Z, H$ denotes $M^2_A, M^2_Z, M^2_H$ respectively. $\delta Z$ means $\delta M^2_Z$ and so on.
Similar situation had ever happened in SM, although the counterterm in our Eq.~(\ref{db}) looks like a little queer. 
The radiative corrections had never shrunk one from defining the counterterm of Weinberg angle $\delta\theta_W$ by $\cos^2\theta_W=M^2_W/M^2_Z$ even before 
the discovery of gauge bosons $W$ and $Z$. 

The renormalization constant of other mixing is get as soon as $M^2_{hP}$ appears,
\begin{eqnarray}
& &\delta Z_{Hh}=2 [~\frac{\Sigma_{Hh} (M^2_{hP})}{M^2_{HP}-M^2_{hP}}+\frac{T_{Hh}}{(M^2_{HP}-M^2_{hR})}+\frac{(M^2_{hP}-M^2_{hR})\Sigma_{Hh} (M^2_{HP})}{(M^2_{HP}-M^2_{hR})(M^2_{HP}-M^2_{hP})}~]\nonumber\\
& &\delta Z_h=-\frac{d}{dq^2} \Sigma_{hh} (q^2)|_{q^2=M^2_{hP}}
\label{Hhmix}
\end{eqnarray}
which is useful for practical manipulation. Similar treatment can be applied to the charged Higgs
as listed in the Appendices. The frequently used rotation angle $\alpha$ of CP even Higgs is defined  
in the first of Eq.~(\ref{v1v2v3}) and its counterterm is 
\begin{equation}
\delta \alpha =\frac{1}{2}\csc^2 2\alpha [~2\csc^2 2\beta \frac{M^2_Z+M^2_A}{M^2_A-M^2_Z}\delta\beta
+ 2\tan 2\beta\frac{M^2_Z \delta M^2_Z - M^2_A \delta M^2_A}{{(M^2_A-M^2_Z)}^2} ~]
\label{da}
\end{equation}
Now, the radiative corrected tree-level relations are no more than Eq.~(\ref{treemass}). 
If we do not insisted that $M^2_{hR}$ ($M^2_{H^+ R}$) is physical mass of (charged) Higgs, even these equations hold. 
The perturbative MSSM permits a chance to calculate the dependence of $M^2_{hP}, M^2_{H^+}, \beta$ and $\alpha$ on 
$ e, M^2_Z, M^2_W, M^2_A$ and $M^2_H$, which will be given as input parameters from experiments.

A numerical investigation is given in the 5'th section, where we will be convinced that, the re-parameterization $\beta\rightarrow M^2_H$ has never caused any considerable numerical distinction from other schemes. And now we turn to 
another important aspect of MSSM renormalization, the consequence of gauge invariance. 

\section{renormalization of gauge-Higgs mixing from WTI}
\setcounter{equation}{0}\setcounter{footnote}{0}

The appropriate renormalization of gauge fixing term (consequently of gauge-Higgs mixing), which should be consistent with the renormalization of VEV (consequently of $\beta$), must be studied carefully here.
Following \cite{ross} and \cite{aoki}, we change nothing else but attach a 
subscription R (meaning renormalized) to the fields and parameters in Eq.~(\ref{zgf}). 
Then in our scheme, the renormalization keeps the form of Eq.~(\ref{zgf}) unchanged at all. 
The main reason is that the renormalization procedure mentioned above on the classical 
Lagrangian has really cancelled all the UV divergence of the proper vertices. 
It's convenient to examine this point by one auxiliary generating functional action \cite{aoki}
\begin{eqnarray}
& &{\bar \Gamma}[F,K]=-i \log \{ \int [{\cal D}F(x)]Exp[i \int dx {\cal L}(x)_{eff}]+J(x) F(x)+K(x) {\delta}^{BRS} F(x) \} \nonumber \\
& &+i \log \{ \int [{\cal D}F(x)]Exp[i \int dz {\cal L}(z)_{gf}\}
\label{nogf}
\end{eqnarray}
where $\int [{\cal D}F(x)] $ denotes the functional integrating of all the fields $F(x) $, such as vector, scalar and ghost fields. 
$K(x)$ is the source of the BRST transformation on field $F(x)$, corresponding term is added to
\begin{equation}
{\cal L}_{eff}={\cal L}_{cl}+{\cal L}_{gf}+{\cal L}_{FP}
\end{equation}
Since the fields power in gauge fixing term is no higher than two, the contribution 
of ${\cal L}_{gf}$ to the proper vertices from ${\bar \Gamma[F,K]}$ is merely in loops. So the 
deduction in Eq.~(\ref{nogf}) is equivalent to that no renormalization substitution is needed 
for ${\cal L}_{gf} $ within ${\cal L}(x)_{eff}$. 
Further expounding necessitates the important WTI which is held in
both 2HDM and MSSM, 
\begin{eqnarray}
&~&\int d^4 x 
\frac{\delta {\bar \Gamma}}{\delta Z_{\nu}(x)}\frac{\delta {\bar \Gamma}}{\delta K_Z^{\nu}(x)}+
\frac{\delta {\bar \Gamma}}{\delta A_{\nu}(x)}\frac{\delta {\bar \Gamma}}{\delta K^{\nu}_\gamma(x)}+\frac{\delta {\bar \Gamma}}{\delta G(x)}\frac{\delta {\bar \Gamma}}{\delta K_G(x)}\nonumber\\
&+&\frac{\delta {\bar \Gamma}}{\delta A(x)}\frac{\delta {\bar \Gamma}}{\delta K_A(x)}+\frac{\delta {\bar \Gamma}}{\delta C^Z(x)}\frac{\delta {\bar \Gamma}}{\delta K_{C^Z}(x)}+
\frac{\delta {\bar \Gamma}}{\delta C^{\gamma}(x)}\frac{\delta {\bar \Gamma}}{\delta K_{C^{\gamma}}(x)}
=0
\label{WT}
\end{eqnarray} 
where the symbolic $F(x)$ has been embodied as the neutral vector boson $Z_{\mu}$, photon 
$A_{\mu}$, their corresponding ghost $C^Z$ ($C^{\gamma}$), the neutral unphysical Goldstone G and 
the pseudo-scalar A. The $K_i(x)$ is the BRST source coupled to corresponding field.
When the functional differentiates
$\frac{{\delta}^2}{\delta Z_{\mu}(y_1) \delta C^Z(y_2)}$, 
$\frac{{\delta}^2}{\delta G(y_1) \delta C^Z(y_2)}$, 
$\frac{{\delta}^2}{\delta A(y_1) \delta C^Z(y_2)}$,
$\frac{{\delta}^2}{\delta A_{\mu}(y_1) \delta C^Z(y_2)}$,
$\frac{{\delta}^2}{\delta A_{\mu}(y_1) \delta C^{\gamma}(y_2)}$
$\frac{{\delta}^2}{\delta A_{\mu}(y_1) \delta C^Z(y_2)}$
are performed on Eq.~(\ref{WT}), a set of WTI as Eqs.~({\ref{ghwt}) are obtained.
There ${\tilde \Gamma}^x_{i,j}$ denotes two points (1PI) vertex in 
momentum space, and ${\tilde \Gamma}[C^i,K_j]$ denotes the Fourier transformation of 
$ \frac{{\delta}^2 {\bar\Gamma}}{\delta C^i \delta K_j}$. The latter can be calculated from
\begin{eqnarray}
{\delta}^{BRS} G &=&\frac{gv}{2} C^Z  - \frac{g_1}{2} [~C^+ G^- + C^- G^+ ~]+\frac{g}{2} C^Z [~H \cos (\alpha-\beta)- h \sin(\alpha-\beta)~]\nonumber\\
{\delta}^{BRS} A &=&\frac{g}{2} C^Z [~h \cos (\alpha-\beta) + H \sin (\alpha-\beta)~]-\frac{g_1}{2} [~C^+ H^- + C^- H^+ ~]\nonumber\\
{\delta}^{BRS} Z_{\mu}&=&-\frac{i g_1^2}{g} (W^+_{\mu} C^- - W^-_{\mu} C^+)+{\partial}_{\mu} C^Z \nonumber \\
{\delta}^{BRS} A_{\mu}&=&-\frac{i g_1 g_2}{g} (W^+_{\mu} C^- - W^-_{\mu} C^+)+{\partial}_{\mu} C^A 
\label{ghbrs}
\end{eqnarray}
for example,
\begin{equation}
{\tilde \Gamma}[C^Z,K_Z^{\nu}]=k_{\nu} J(k^2),~~~~{\tilde \Gamma}[C^Z,K_G]= -i M_Z I(k^2)
\end{equation}
and so on. It's luck to find that, for most of those physical vertices ${\tilde \Gamma}^x_{i,j}$, their coefficient functions
 usually vanish at tree-level. Even though those coefficient might remain, these unphysical vertices ${\tilde \Gamma}[C^i,K_j]$ can 
be eliminated away as the treatment in \cite{aoki}. Furthermore, when only one-loop corrections of ${\tilde \Gamma}^x_{i,j}$ 
are concerned, ${\tilde \Gamma}[C^i,K_j]$ can be kept at lower order, then the tree form  $J(k^2)=1,~I(k^2)=1$ is sufficient for these equations,
\begin{eqnarray}
& &{\tilde \Gamma}^{\mu\nu}_{ZZ} k_\nu + {\tilde \Gamma}^{\mu}_{ZG} (-i M_Z)=0,~~~
{\tilde \Gamma}^{\nu}_{GZ} k_\nu + {\tilde \Gamma}_{GG}(-i M_Z) =0
\nonumber \\
& &{\tilde \Gamma}^{\nu}_{AZ} k_\nu +{\tilde \Gamma}_{AG} (-i M_Z)=0,~~~
{\tilde \Gamma}^{\mu\nu}_{\gamma Z} k_\nu +{\tilde \Gamma}^{\mu}_{\gamma G} (-i M_Z)=0 
\nonumber \\
& &{\tilde \Gamma}^{\mu\nu}_{\gamma Z} k_\nu =0,~~~
{\tilde \Gamma}^{\nu}_{A \gamma} k_\nu =0 
\label{ssghwt}
\end{eqnarray}
Except the third and the sixth, these equations are recognized just as the ones met in SM. For example, 
\cite{hollik} had given similar expressions deduced from the generating functional of the full Green function. 

With the definition of the renormalization constants $Z_{ZG}, \delta Z_{AZ}, \delta Z_{\gamma G}$ and $\delta Z_{A\gamma}$,
\begin{eqnarray}
& &{{\tilde \Gamma}^{R~~\mu}}_{ZG}=Z_{ZG}(-i M_Z k^{\mu})+ {\tilde \Gamma}^{\mu}_{ZG},~~~
{{\tilde \Gamma}^{R~~\mu}}_{GZ}=Z_{ZG}(i M_Z k^{\mu})+ {\tilde \Gamma}^{\mu}_{GZ} \nonumber \\
& &{{\tilde \Gamma}^{R~~\mu}}_{AZ}=\delta Z_{AZ}(i M_Z k^{\mu})+ {\tilde \Gamma}^{\mu}_{AZ},~~~
{{\tilde \Gamma}^{R~~\mu}}_{\gamma G}=\delta Z_{\gamma G} (-i M_Z k^{\mu})+{\tilde \Gamma}^{\mu}_{\gamma G} \nonumber \\
& &{{\tilde \Gamma}^{R~~\mu}}_{A\gamma}=\delta Z_{A\gamma} (i M_Z k^{\mu})+{\tilde \Gamma}^{\mu}_{A\gamma} 
\label{wt}
\end{eqnarray}
those equations in (\ref{ssghwt}) can constrain the renormalized vertices too,
\begin{eqnarray}
& &k_{\nu} [~(M^2_Z+\delta M^2_Z) Z_Z \frac{k^{\mu}k^{\nu}}{k^2} + {\tilde \Gamma}^{\mu\nu}_{ZZ}~]
+(-i M_Z) [~Z_{ZG}(-i M_Z k^{\nu})+ {\tilde \Gamma}^{\mu}_{ZG}~]=0 \nonumber\\
& &k_{\nu}[~Z_{ZG}(i M_Z k^{\nu})+ {\tilde \Gamma}^{\mu}_{GZ}~]
+(-i M_Z)[~Z_G k^2 +{\tilde \Gamma}_{GG}~]=0 \nonumber\\
& &k_{\nu}[~\delta Z_{ZA}(i M_Z k^{\nu})+ {\tilde \Gamma}^{\mu}_{AZ}~]
+(-i M_Z)[~ Z^{1/2}_{GA} M^2_A +{\tilde \Gamma}_{AG}~]=0 \nonumber\\
& &k_{\nu} [~(M^2_Z+\delta M^2_Z) Z^{1/2}_Z Z^{1/2}_{Z\gamma}\frac{k^{\mu}k^{\nu}}{k^2} 
+ {\tilde \Gamma}^{\mu\nu}_{Z\gamma}~]+(-i M_Z) [~\delta Z_{\gamma G}(-i M_Z k^{\nu})+ {\tilde \Gamma}^{\mu}_{\gamma G}~]=0 \nonumber\\
& &k_{\nu} [~(M^2_Z+\delta M^2_Z) Z_{Z\gamma}\frac{k^{\mu}k^{\nu}}{k^2} + {\tilde \Gamma}^{\mu\nu}_{\gamma Z}~]=0,~~~
k_{\nu} [\delta Z_{A\gamma} (i M_Z k^{\nu})+{\tilde \Gamma}^{\mu}_{A\gamma} ~]=0
\label{loopwt}
\end{eqnarray}
$Z_{ZG}$ comes out from Eq.~(\ref{loopwt}) up to one loop order, when the first contracted with $k_{\mu}$ is added to the second produced with $i M_Z$.
The last two are the trivial constrains which repeat the fact that, the propagator of massless photon is transverse and $A^{\mu}-A$ mixing is UV convergent at one-loop.
\begin{eqnarray}
& &\delta Z_{ZG}=\frac{1}{2} \delta Z_Z +\frac{1}{2} \delta Z_G + \frac{\delta M^2_Z}{2 M^2_Z} \nonumber\\
& &\delta Z_{AZ}=\frac{1}{2}\delta Z_{GA}~~~
\delta Z_{\gamma G}=\frac{1}{2}\delta Z_{Z\gamma}~~~
\delta Z_{\gamma A}=0
\label{mixga}
\end{eqnarray}

It is the full proper vertex (with gauge fixing term) that goes into physical calculation. 
When the gauge fixing term in Eq.~(\ref{nogf}) is restored, only the unit term in the original gauge-scalar mixing are 
cancelled, since the gauge fixing terms are kept unchanged in this scheme. Then such scheme gives
\begin{eqnarray}
{\Gamma}^{R~~\mu}_{ZG}=\delta Z_{ZG}(-i M_Z k^{\nu})+ {\tilde \Gamma}^{\mu}_{ZG} \nonumber\\
{\Gamma}^{R~~\mu}_{AZ}={\tilde \Gamma}^{R~~\mu}_{AZ},~~~{\Gamma}^{R~~\mu}_{\gamma G}={\tilde \Gamma}^{R~~\mu}_{\gamma G}
\label{ZG}
\end{eqnarray}
It's worthy to notice that, the $\delta {\cal M}_{AG}$ (tadpole in Eq.~(\ref{AAGcoun})) must be included into the $\tilde\Gamma_{AG}$. 
Otherwise, the UV divergence in the $Z_\mu -G$ transition can not be cancelled neatly as formally imaged by the third of Eq.~(\ref{ssghwt}). 
Since these renormalization constants of mixing ought to be inserted into the 
amplitude of some physical process, these tadpoles mustn't be dropped away naively, and they support the $\delta\beta$ for a finite S-Matrix. 
In such a way, when the relations of proper vertices are constructed appropriately, no doubt, Eq.~(\ref{ssghwt}) guarantees 
that such a perturbative definition can make those renormalized Green functions on the left side of Eq.~(\ref{wt}) to be UV convergent. 

For a comparing, we perform the renormalization replacement to the linear order of $\delta$ listed in Eq.~(\ref{wave}) but
leave ${\cal L}_{gf} $ unchanged, and collect the dimensionless coefficients of the gauge-scalar fixings $Z_{\mu} {\partial}^{\mu} G$,
 $Z_{\mu} {\partial}^{\mu} A$, $A_{\mu} {\partial}^{\mu} G$ and $A_{\mu} {\partial}^{\mu} A$ respectively, 
 then we find ourselves run onto Eq.~(\ref{mixga}) again. 
\begin{eqnarray}
& &{\cal L}_{mix}= -M_Z Z^{\mu} \partial_{\mu} G  \rightarrow \nonumber\\
& &-( M_Z + \delta M_Z)[~Z^{1/2}_Z Z^{\mu}+Z^{1/2}_{Z\gamma} A^{\mu}~] ~\partial_{\mu}~ [~Z^{1/2}_{GA} A+ Z^{1/2}_G G~]\nonumber\\
&=&-M_Z[~\frac{1}{2} \delta Z_Z +\frac{1}{2} \delta Z_G + \frac{\delta M_Z}{M_Z}~] ~~Z^{\mu} \partial_{\mu} G \nonumber\\
& &-M_Z[~\frac{1}{2} \delta Z_{GA}~] ~~Z^{\mu} \partial_{\mu} A \nonumber\\
& &-M_Z[~\frac{1}{2} \delta Z_{Z\gamma} ~] ~~A^{\mu} \partial_{\mu} G \nonumber\\
& &+0~~ A^{\mu} \partial_{\mu} A
\end{eqnarray}
The similar situation take place for electric charged gauge-scalar mixing and all other 
truncated Green functions. This means nothing else but the WTI has ensured 
the renormalization ``leaving ${\cal L}_{gf} $ out" can cancel all these UV divergence.

In fact we could have renormalized ${\cal L}_{gf} $ with explicit renormalization constants like the last line of Eq.~(\ref{gaupara}). 
In consequence, we would have to seek proper counterterms for $\alpha_z$ \cite{hollik}, 
so that all these divergence can be cancelled up within the ${\cal L}_{gf} $ terms.
Then we succeed the inference of the work of \cite{bwlee} to economize the renormalization 
of ${\cal L}_{gf} $ at the beginning. 

\section{ discussions and conclusions }
When such a systematic renormalization scheme is completed of Higgs sector and gauge-scalar mixing, the calculation of S-matrix 
can be organized in an apparent and simple way like \cite{denner}and \cite{pierce}.
Here we concern only that, by which feature and to what extent can we judge the lightest Higgs is 
supersymmetric or not, if it is awaken up here or there. So  
Eqs. ~(\ref{hpolemass}, ~\ref{Hhmix}, ~\ref{db}, ~\ref{da}) have to be employed for a complete simulation. The Eq.~(\ref{treemass}) set the mass of the lightest and the charged Higgs when they appear in the inner line of loops as virtual particles.
A symbolic $M^2_{hP} $ can save bookkeeping and these expressions are very simple for modern computers although they seem tedious. 

After the Taylor expanding in Eq.(\ref{hpolemass}), the manifestation of the 
UV cancelation is straightforward in our analysis expressions, with the help from $Cos2\alpha = -Cos2\beta (M^2_A-M^2_Z) / (M^2_H-M^2_{hR} )$, ~
$ M^2_{hR}+M^2_H=M^2_A+M^2_Z $ ~ and a relation in the stop sector 
$2 M_t(A_u +\mu Cot\beta )=(M^2_{{\tilde t}_1}-M^2_{{\tilde t}_2}) Sin2\theta_{\tilde t}$. ($\theta_{\tilde t}$ is the mixing angle between the left and the right hand stop quarks). However that series is not convenient for a numerical solution, so we iterate $q^2$ near $M^2_{hR}$ until our equation is satisfied with a standard $FF$ package \cite{ff}.
 In our scheme the expression (\ref{sequen}) means that we can not make a global plot for $M_{hP}$ dependent on $ M_A $ ($M_H$) when $M_H$ ($M_A$) is fixed. 
We investigated the range $ M_A (M_H) \sim 110, 250, 500, 850 ~GeV $ respectively with a top quark mass $M_t=175$ GeV as shown in the Fig.1.
As to the parameters in the stop quark sector, we prefer to the physical masses of stops and their mixing angle. 
A set of representative inputs are plotted, such as {\it light} spectra just over experiment bound \cite{stops} ( $M_{{\tilde t}_1}=70$~ $M_{{\tilde t}_2}=230$ ) and a {\it heavy} spectra ( $M_{{\tilde t}_1}=250$~ $M_{{\tilde t} _2}=850$ GeV), 
with both zero ($\theta_{\tilde t}=0$) and maximal ($\theta_{\tilde t}=\pi /4$) mixing.
One can still recognize $M_{hP}$ from the profile in our figures, although they seem a little unfamiliar to the eyes accustomed to conventional $tan\beta$ plots. 
The data displayed are merely by-production of our scheme, from which we
can conclude that the dominant radiative corrections for the lightest Higgs has been acquainted properly,
although a more accurate prediction for its mass is not reached since neither the whole virtual particles nor the two loop effect were included in our numerical reiteration.

This FDC procedure is designed for a simple and consistent amplitudes calculation without EP or RG, since less junction means less uncertainty. 
This realization is also compatible with taking over the conventional treatment in \cite{aoki,hollik,denner,santos} 
for the renormalization of SM gauge bosons, fermions and couplings.

This systematic renormalization for MSSM need the super partners of the involved virtual particles, no matter how heavy they would be. The Decoupling Theorem \cite{decouple} still holds but in a manner that particles have to decouple with 
their corresponding super partner. Otherwise the UV divergence would be left 
in mass of the light Higgs boson. 
Using Decoupling Theorem in the manner that all the super partners are integrated out from the original Lagrangian, means the scheme for 2HDM in \cite{santos}, 
then the price is that the all the masses of Higgs bosons and the angles $\beta$ , $\alpha$ have to be input as FDC free parameters.
 
Anyway, that 2HDM and this MSSM have the same gauge structure, so our Eqs.~(\ref{mixga},~\ref{ZG}) is still held and utilizable. 
Certainly, some decay of Higgs can be employed to renormalize $\beta, \alpha$ in this unconstrained 2HDM.

{\it Acknowledgements}

The author would like to thank Prof. Zhang Zhao-Xi. This work was supported by National 
Natural Science Foundation of China.

\appendix
\section*{}
\renewcommand{\thesubsection}{\Alph{subsection}}
\renewcommand{\theequation}{\Alph{subsection}.\arabic{equation}}

\subsection{ mass v.s. gauge eigenstates in MSSM}
\setcounter{equation}{0}
A. MSSM field representation

\begin{eqnarray}
  \left(\matrix{H\cr h}\right),~~\left(\matrix{G \cr A}\right),~~\left(\matrix{G^+ \cr H^+}\right)
  =\left(\matrix{\cos\zeta & \sin\zeta
              \cr - \sin\zeta & \cos\zeta }\right)
  \left(\matrix{\phi^0_1 \cr \phi^0_2 }\right),~~\left(\matrix{\chi^0_1 \cr \chi^0_2 }\right),~~\left(\matrix{\phi^+_1 \cr \phi^+_2 }\right)
\label{angles}
\end{eqnarray}
where $\zeta= \alpha, ~\vartheta,~\vartheta_+$ from left to right respectively.
At tree-level $ \vartheta=\vartheta_+=\beta$ and this renormalization can accommodate 
$ \vartheta^R=\vartheta^R_+=\beta^R$ and $\delta\vartheta=\delta\vartheta_+=\delta\beta$. 

B. $m^2_i$ expressed as physical parameters

$m^2_1,~m^2_2,~m^2_3$ can be solved as function of ${\cal M}_{AA}, ~{\cal T}_H,~{\cal T}_h$ in MSSM
\begin{eqnarray}
m^2_1&=&-\sec^2 (\beta - \vartheta) \{ [-16 {\cal M}_{AA} + 2 (8 {\cal M}_{AA} + 2 g^2 v^2 \cos^2 (\beta - \vartheta)) \cos 2\beta ] /32\nonumber\\
&-&{\cal T}_H\cos \vartheta [ \cos ( \alpha  - \beta - \vartheta)  - 2 \cos ( \alpha  + \beta - \vartheta)  -  \cos ( \alpha  - \beta + \vartheta) ]/(2 v)\nonumber\\
&-&{\cal T}_h \cos \vartheta [-\sin ( \alpha  - \beta - \vartheta)  + 2 \sin ( \alpha  + \beta - \vartheta)  + \sin ( \alpha  - \beta + \vartheta) ]/(2 v) \}\nonumber\\
m^2_2&=&\sec^2 (\beta - \vartheta) \{ [16 {\cal M}_{AA} + 2 (8 {\cal M}_{AA} +2 g^2 v^2 \cos^2 (\beta - \vartheta)) \cos 2\beta ]/32\nonumber\\
&+&{\cal T}_h \sin \vartheta [\cos ( \alpha  - \beta - \vartheta)  + 2 \cos ( \alpha  + \beta - \vartheta)  + \cos ( \alpha  - \beta + \vartheta) ]/(2 v)\nonumber\\
&+&{\cal T}_H \sin \vartheta [\sin ( \alpha  - \beta - \vartheta)  + 2 \sin ( \alpha  + \beta - \vartheta)  + \sin ( \alpha  - \beta + \vartheta) ]/(2 v) \}\nonumber\\
m^2_3&=&\sec^2 (\beta - \vartheta) \{{\cal T}_h [\cos ( \alpha  + \beta)  + \cos ( \alpha  - \beta)  \cos 2\vartheta ]/(2 v) \nonumber\\
&+&{\cal T}_H [\cos 2\vartheta \sin ( \alpha  - \beta)  + \sin ( \alpha  + \beta) ]/(2 v)\nonumber\\
&-& {\cal M}^2_A \sin 2\beta /2 \}
\label{ms}
\end{eqnarray}

C. Mass form 

Multiplied by a factor 2, the matrix elements rotated from the original potential read,
\begin{eqnarray}
{\cal M}_{AG}&=& {\cal M}_{GA}=[~8 m^2_3 \cos 2\vartheta  + (-4 m^2_1 + 4 m^2_2 - g_1^2 v_1^2 - g_2^2 v_1^2 + g_1^2 v_2^2 + 
       g_2^2 v_2^2) \sin 2\vartheta ~]/4\nonumber\\
{\cal M}_{GG}&=&[~(4 m^2_1 - 4 m^2_2 + g_1^2 v_1^2 + g_2^2 v_1^2 - g_1^2 v_2^2 - g_2^2 v_2^2) \cos 2\vartheta  + 
    4 (m^2_1 + m^2_2 + 2 m^2_3 \sin 2\vartheta )~]/8\nonumber\\
{\cal M}_{hh}&=&[~4 m^2_1 + 4 m^2_2 + g_1^2 v_1^2 + g_2^2 v_1^2 + g_1^2 v_2^2 + g_2^2 v_2^2 \nonumber\\
& &- 2 (2 m^2_1 - 2 m^2_2 + g_1^2 v_1^2 + g_2^2 v_1^2 - g_1^2 v_2^2 - g_2^2 v_2^2)\cos 2\alpha  \nonumber\\
& &+ 2 (-4 m^2_3 + g_1^2 v_1 v_2 + g_2^2 v_1 v_2) \sin 2\alpha ~]/8\nonumber\\
{\cal M}_{hH}&=&{\cal M}_{Hh}=[~(4 m^2_3 - (g_1^2 + g_2^2) v_1 v_2) \cos 2\alpha  \nonumber\\
& &+ (-2 m^2_1 + 2 m^2_2 - g_1^2 v_1^2 - g_2^2 v_1^2 + g_1^2 v_2^2 + g_2^2 v_2^2) \sin 2\alpha ~]/2\nonumber\\
{\cal M}_{H^+ H^-}&=&(4 m^2_1 + 4 m^2_2 + g_2^2 v_1^2 + g_2^2 v_2^2 \nonumber\\
& &+ (-4 m^2_1 + 4 m^2_2 - g_1^2 v_1^2 + g_1^2 v_2^2) \cos 2\vartheta_+  \nonumber\\
& &+ 2 (-4 m^2_3 + g_2^2 v_1 v_2) \sin 2\vartheta_+ ~]/4\nonumber\\
{\cal M}_{H^+ G^-}&=&{\cal M}_{G^+ H^-}=[~2 (4 m^2_3 - g_2^2 v_1 v_2) \cos 2\vartheta_+  + 
    (-4 m^2_1 + 4 m^2_2 - g_1^2 v_1^2 + g_1^2 v_2^2) \sin 2\vartheta_+ ~]/4\nonumber\\
{\cal M}_{G^+ G^-}&=&(4 m^2_1 + 4 m^2_2 + g_2^2 v_1^2 + g_2^2 v_2^2 \nonumber\\
& &+ (4 m^2_1 - 4 m^2_2 + g_1^2 v_1^2 - g_1^2 v_2^2) \cos 2\vartheta_+ \nonumber\\
& &+ 2 (4 m^2_3 - g_2^2 v_1 v_2) \sin 2\vartheta_+ ~]/4
\end{eqnarray}

\subsection{counterterms in MSSM }
\setcounter{equation}{0}
A. The counterterms for the combinations of tadpoles are
\begin{eqnarray}
& &T_{AA}=0 \nonumber\\
& &T_{HH}=\frac{-1}{2 v}\cos (\alpha-\beta) [~-3 T_H  + T_H  \cos 2(\alpha-\beta)  - T_h  \sin 2(\alpha-\beta) ~]
\nonumber\\
& &T_{AG}=T_{GA}=\frac{2 }{v} [~T_h  \cos (\alpha-\beta)  + T_H  \sin (\alpha-\beta) ~]
\nonumber\\
& &T_{GG}=\frac{1 }{v}[~T_H  \cos (\alpha-\beta)  - T_h  \sin (\alpha-\beta) ~]
\nonumber\\
& &T_{hh}=\frac{-1}{2 v}~\sin (\alpha-\beta)  [~3 T_h  + T_h  \cos 2(\alpha-\beta)  + T_H  \sin 2(\alpha-\beta) ~]
\nonumber\\
& &T_{Hh}=T_{hH}=\frac{1}{2 v}[~3 T_h  \cos (\alpha-\beta)  + T_h  \cos 3(\alpha-\beta)  - 4 T_H  \sin^3 (\alpha-\beta) ~]
\nonumber\\
& &T_{H^+ H^-}=0\nonumber\\
& &T_{H^+ G^-}=T_{G^+ H^-}=\frac{2 }{v} [~T_h  \cos (\alpha-\beta)  + T_H  \sin (\alpha-\beta) ~]
\nonumber\\
& &T_{G^+ G^-}=\frac{2 }{v}[~T_H  \cos (\alpha-\beta)  - T_h  \sin (\alpha-\beta) ~]
\end{eqnarray}
These tadpoles correction in CP even neutral Higgs are different from the ones get by others \cite{pierce,santos}. It's easy to check that our $T_{HH} + T_{hh}
$ is equal to the $b_{HH} + b_{hh} - b_{AA} $ in \cite{pierce}, and the $a_{GG}$ in \cite{gunion}.
However, the Goldstone-Higgs mixing terms are the same, 
as pointed in the context, these Goldstone-Higgs tadpole loops must be included into the corresponding proper vertex.

B. renormalization for charged Higgs sector

 Pole mass and $H^+ G^+$ mixing to one loop order
\begin{eqnarray}
&~&\Sigma_{H^+} (q^2)+ (q^2-M^2_{H^+ R}) Z_{H^+}- \delta M_{H^+}^2= 0\nonumber\\
&~&\delta M^2_{H^+}=\delta M^2_A + \delta M^2_Z\nonumber\\
&~&\Sigma_{H^+G^+} (0)+ (0-M^2_{H^+ R}- \delta M_{H^+}^2) Z^{1/2}_{H^+} Z_{H^+G^+}^{1/2}-T_{G^+H^+} =0\nonumber\\
&~&\Sigma_{H^+G^+} (q^2)+ (q^2-M^2_{H^+ R}- \delta M_{H^+}^2) Z^{1/2}_{H^+} Z_{H^+G^+}^{1/2}+ Z_{G^+G^+}^{1/2} Z_{G^+H^+}^{1/2} q^2 - T_{G^+H^+}= 0\nonumber\\
&~&\frac{d}{dq^2} \Sigma_{H^+} (q^2) + Z_{H^+} + Z_{H^+G^+} = 1 
\label{charH}
\end{eqnarray}
The third equation restricts the charged Goldstone pole mass to be zero, and it's helpful to solve 
$\delta Z_{G^+H^+}$ in the fourth equation, which is more useful for physical process.
There is analogous expression for the neutral Goldstone but the $\delta Z_{GA}$ can be calculated independently.
 
\subsection{Some Ward-Takahashi identities in the neutral sector of MSSM}
\setcounter{equation}{0}

\begin{eqnarray}
{\tilde \Gamma}^{\mu\nu}_{ZZ} {\tilde \Gamma}[C^Z,K_Z^{\nu}]+
{\tilde \Gamma}^{\mu\nu}_{Z\gamma} {\tilde \Gamma}[C^Z,K^{\nu}_\gamma]+
{\tilde \Gamma}^{\mu}_{ZG} {\tilde \Gamma}[C^Z,K_G]+
{\tilde \Gamma}^{\mu}_{ZA} {\tilde \Gamma}[C^Z,K_A]=0 
\nonumber \\ 
{\tilde \Gamma}^{\nu}_{GZ} {\tilde \Gamma}[C^Z,K_Z^{\nu}]+
{\tilde \Gamma}^{\nu}_{G\gamma} {\tilde \Gamma}[C^Z,K^{\nu}_\gamma]+
{\tilde \Gamma}_{GG} {\tilde \Gamma}[C^Z,K_G]+
{\tilde \Gamma}_{GA} {\tilde \Gamma}[C^Z,K_A]=0
\nonumber \\
{\tilde \Gamma}^{\nu}_{AZ} {\tilde \Gamma}[C^Z,K_Z^{\nu}]+
{\tilde \Gamma}^{\nu}_{A\gamma} {\tilde \Gamma}[C^Z,K^{\nu}_\gamma]+
{\tilde \Gamma}_{AG} {\tilde \Gamma}[C^Z,K_G]+
{\tilde \Gamma}_{AA} {\tilde \Gamma}[C^Z,K_A]=0
\nonumber \\
{\tilde \Gamma}^{\mu\nu}_{\gamma Z} {\tilde \Gamma}[C^Z,K_Z^{\nu}]+
{\tilde \Gamma}^{\mu\nu}_{\gamma\gamma} {\tilde \Gamma}[C^Z,K^{\nu}_\gamma]+
{\tilde \Gamma}^{\mu}_{\gamma G} {\tilde \Gamma}[C^Z,K_G]+
{\tilde \Gamma}^{\mu}_{\gamma A} {\tilde \Gamma}[C^Z,K_A]=0 
\nonumber \\
{\tilde \Gamma}^{\mu\nu}_{\gamma Z} {\tilde \Gamma}[C^{\gamma},K_Z^{\nu}]+
{\tilde \Gamma}^{\mu\nu}_{\gamma\gamma} {\tilde \Gamma}[C^{\gamma},K^{\nu}_\gamma]+
{\tilde \Gamma}^{\mu}_{\gamma G} {\tilde \Gamma}[C^{\gamma},K_G]+
{\tilde \Gamma}^{\mu}_{\gamma A} {\tilde \Gamma}[C^{\gamma},K_A]=0 
\nonumber \\
{\tilde \Gamma}^{\nu}_{AZ} {\tilde \Gamma}[C^{\gamma},K_Z^{\nu}]+
{\tilde \Gamma}^{\nu}_{A \gamma} {\tilde \Gamma}[C^{\gamma},K^{\nu}_\gamma]+
{\tilde \Gamma}_{AA} {\tilde \Gamma}[C^{\gamma},K_G]+
{\tilde \Gamma}_{AA} {\tilde \Gamma}[C^{\gamma},K_A]=0 
\label{ghwt}
\end{eqnarray}

The similar expressions hold for the charged sector and lead to the W mixing with scalars,
\begin{eqnarray}
& &\delta Z_{W^+ G^+}=\frac{1}{2} \delta Z_W +\frac{1}{2} \delta Z_{G^+} + \frac{\delta M^2_W}{2 M^2_W} \nonumber\\
& &\delta Z_{W^+ H^+}=\frac{1}{2}\delta Z_{G^+ H^+}~~~
\end{eqnarray}

\begin{figure}[]
\epsfxsize=\columnwidth
\centerline{\epsffile{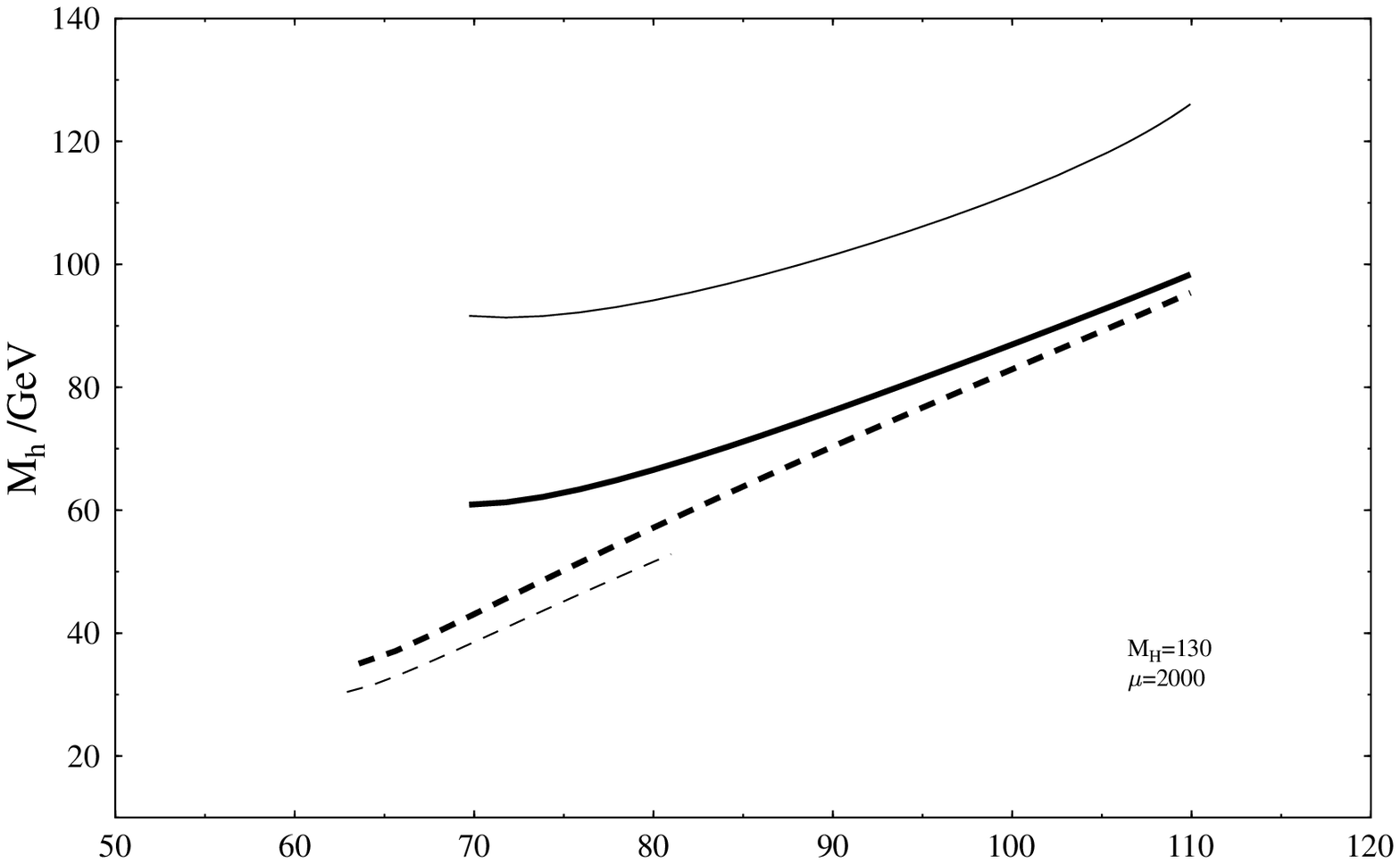}}
\end{figure}

\begin{figure}[]
\epsfxsize=\columnwidth
\centerline{\epsffile{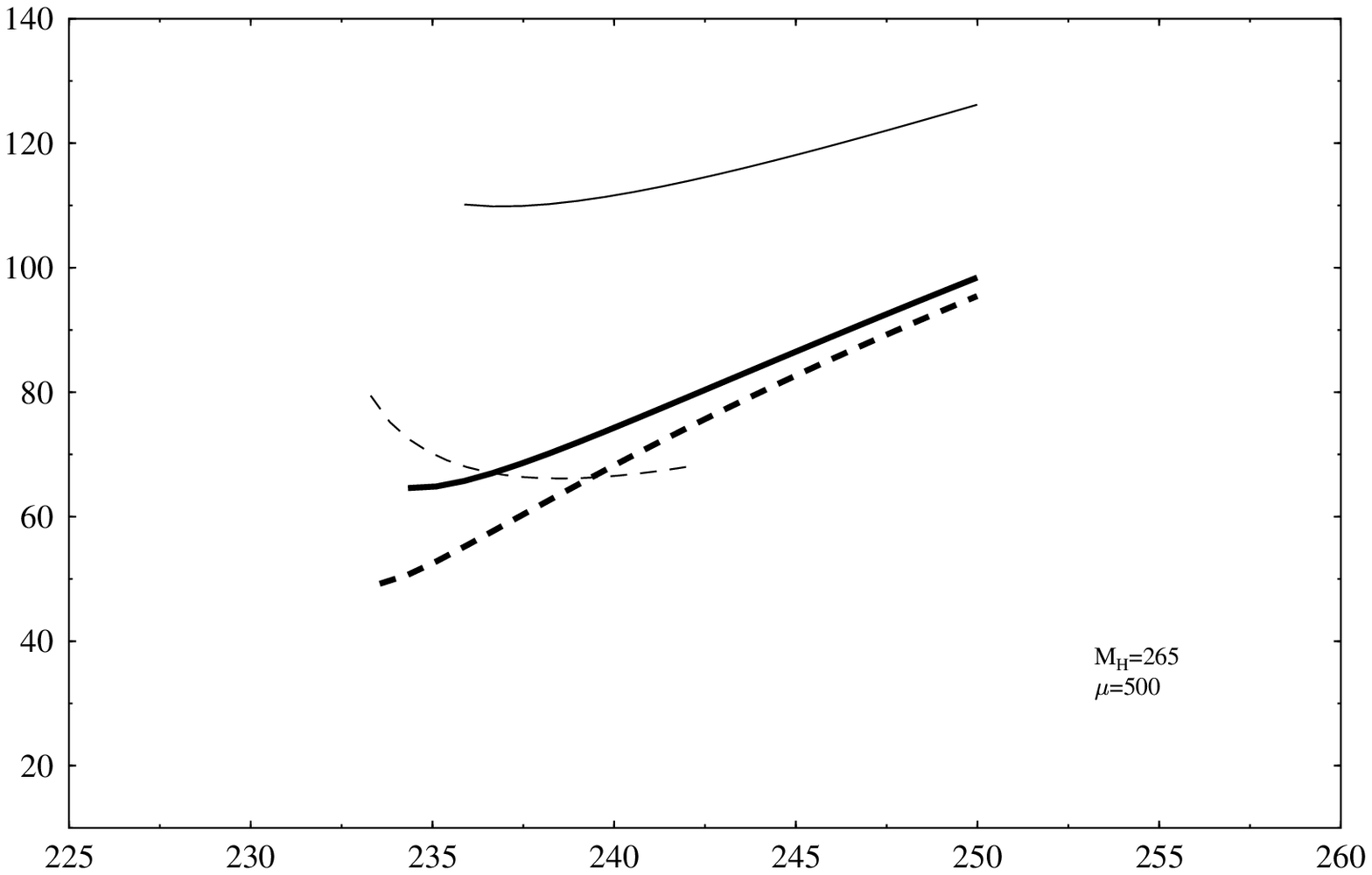}}
\end{figure}

\begin{figure}[]
\epsfxsize=\columnwidth
\centerline{\epsffile{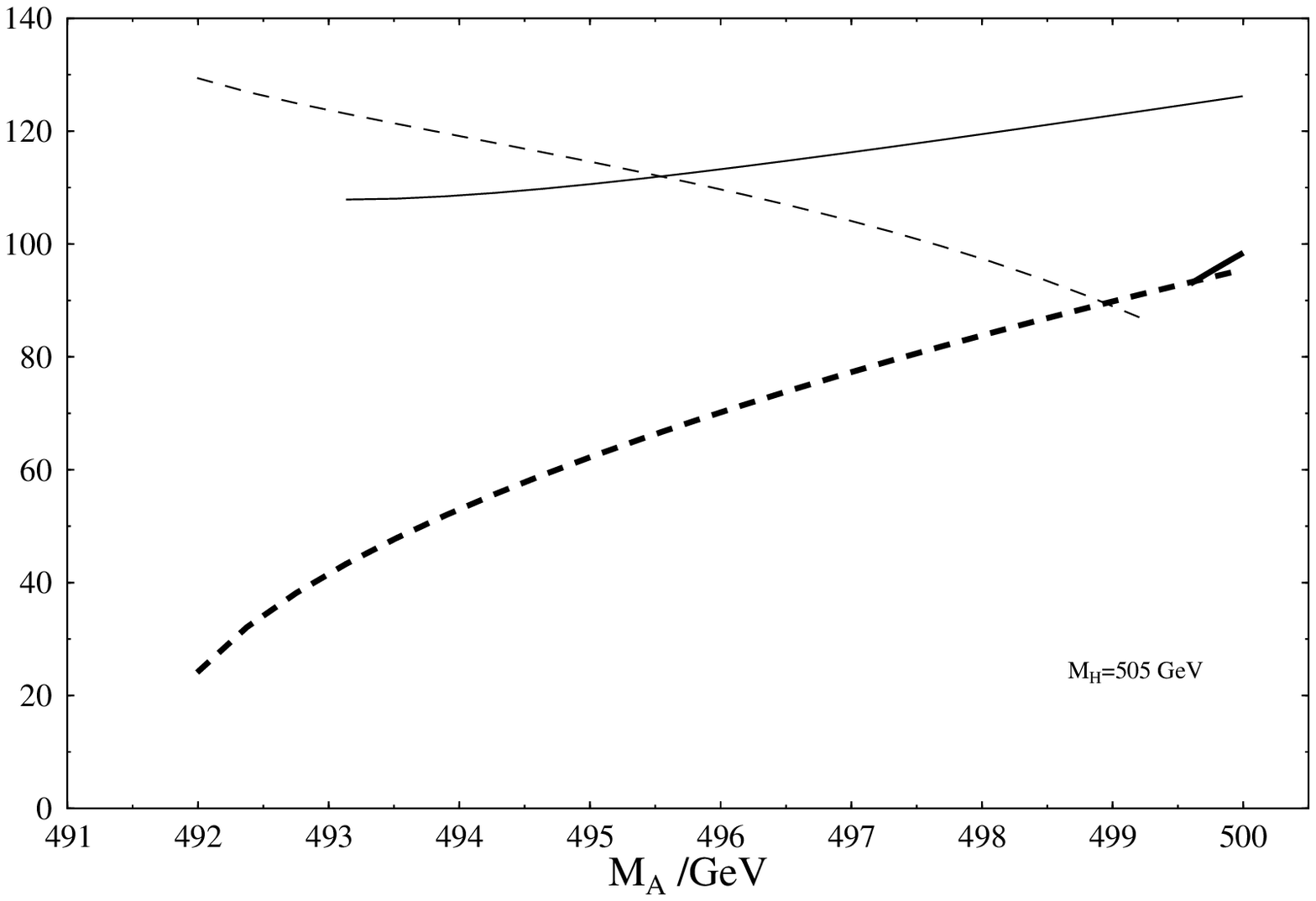}}
\end{figure}

\begin{figure}[]
\epsfxsize=\columnwidth
\centerline{\epsffile{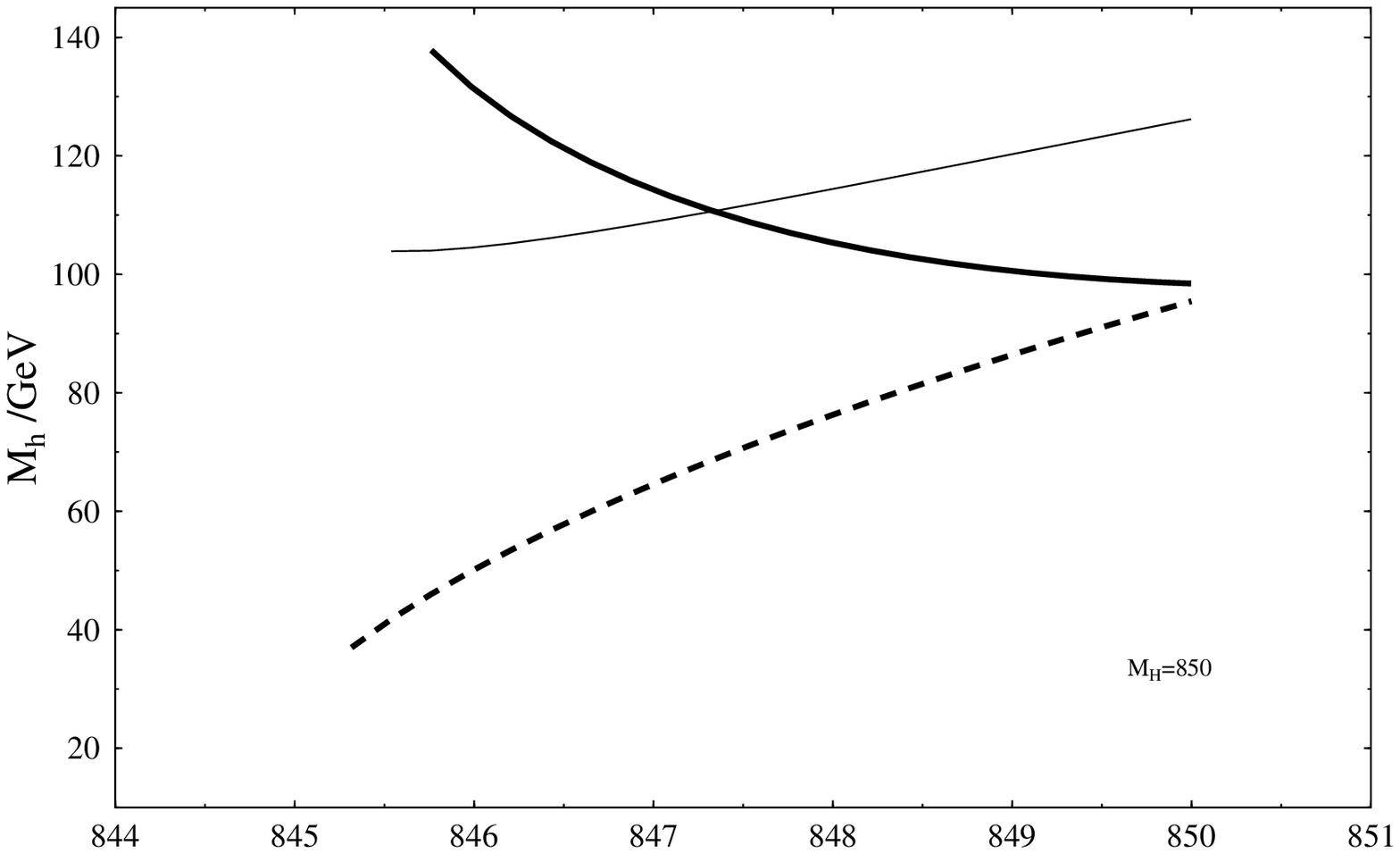}}
\caption{radiatively corrected light CP even Higgs mass is plotted 
as a function of $M_A$, $tan\beta$ varying implicitly from 1.4 (low $M_A$) 
to 80 (heigh $M_A$) with $M_H$ fixed. 
the solid (dashed) lines are for the {\it heavy} ({\it light}) stops, with $\mu =0$ (deferent $\mu $ marked for $\theta_{\tilde t}=0$), and the thin (thick) lines are for zero (maximal) mixing.} 
\end{figure}


\begin{thebibliography}{xxx}
\bibitem{mssm}H.E.Haber and J.L.Kane, Phys. Rep. 117 (1985) 75; C.Csaki, Mod. Phys. Lett. A11 (1996) 599;
 A.Dawson, hep-ph/9712464
\bibitem{guide}J.F.Gunion,H.E.Haber, The Higgs Hunter's Guide (Addison-Wesley, Redwood City, CA, 1990)
\bibitem{ep}J.Ellis et al., Phys. Lett. B257 (1991) 83,Phys. Lett. B262 (1991) 477;
 A.Brignole et al., Phys. Lett. B271 (1991) 123
\bibitem{rg}M.Carena et al., Nucl. Phys. B381 (1992) 66; R.Hempfling,A.Hoang, Phys. Lett. B331 (1994) 99
\bibitem{spira}A. Krause et al., Nucl. Phys. B519 (1998) 73;
\bibitem{colli}Elzbieta Richter-Was, D. Froidevaux, Z. Phys. C76 (1997) 665; 
M.Spira, hep-ph/9711407; H.E.Haber, Int. J. Mod. Phys. A13 (1998) 2263; 
A.V. Gladyshev et al., Nucl. Phys. B498 (1997) 3
\bibitem{phC}P.H.Chankowski et al., Nucl. Phys. B423 (1994) 437, Phys. Lett. B286 (1992) 307
\bibitem{aDa}A.Dabelstein et al., Z. Phys. C. 67 (1995) 495
\bibitem{hollik}M.Bohm et al., Fortschr. Phys. 34 (1986) 687
\bibitem{vdre}A.Dabelstein Nucl. Phys. B456 (1995) 25
V. Driesen, W. Hollik, J. Rosiek, Z. Phys. C71 (1996) 259
\bibitem{SH}S.Heinemeyer et al., Phys. Rev. Lett. 78 (1997) 3626, hep-ph/9803277
\bibitem{aoki}K.I.Aoki, et al., Suppl. Prog. Theor. Phys. 73 (1982) 1
\bibitem{denner}A.Denner, Fortschr.Phys. 41 (1993) 307
\bibitem{santos}R.Santos, Phys. Rev. D56(1996) 5366 
\bibitem{pierce}D.Pierce and A.Papadopoulos, Phys. Rev. D47(1993) 222
\bibitem{kane} H. E. Habber, in: Perspectives on Higgs Physics, Advanced Series on Directions in Heigh energy 
physics, Vol 13, Ed G.L.Kane, (World Scientific, Singapore, 1993)
\bibitem{haber}H.E.Haber, in: The Standard Model and Beyond, The 9'th Symposium on Theoretical Physics, 
Ed J.E. Kim,  (World Scientific, Singapore, 1991)
\bibitem{gunion}J.F.Gunion, A.Turski, Phys. Rev.D 39,2701(1989)
\bibitem{berger}M.S.Berger, Phys. Rev. D41(1990)225 
\bibitem{sola}J.Sola, hep-ph/9712491 
\bibitem{lep}ALEPH Collaboration, Phys. Lett. B412 (1997) 173; OPAL Collaboration (K. Ackerstaff et al.). CERN-EP-98-029; 
L3 Collaboration (M. Acciarri et al.). CERN-EP-98-052
\bibitem{fdc}P.H.Chankowski, et al., Phys. Lett. B274 (1992) 191;
A.Brignole, Phys. Lett. B281(1992) 284
\bibitem{apprada} H.E.Haber, R.Hempfling,and H.Hoang, Z. Phys. C75 (1997) 539
\bibitem{effpoten} M.Carena,M.Quiros,C.E.M.Wagner, Nucl. Phys. B461 (1996) 407
\bibitem{epslon}A.Djouadi et al., Z. Phys. C70 (1996) 435
\bibitem{ross}D.A.Ross and J.C.Taylor, Nucl. Phys. B51 (1973) 125
\bibitem{bwlee}Benjamin W.Lee, Phys. Rev. D9 (1974) 933
\bibitem{ff}G.J.van Oldenborgh and J.A.M.Vermaseren, Z.Phys. C46(1990)425
\bibitem{stops}Particle Data Group, Eur. Phys. Jour. C3 (1998) 769
\bibitem{decouple}B.Ovrut and H.Schnitzer, Phys. Rev. D22 (1980) 2518;
 T.Appelquist and C.Bernard, Phys. Rev. D23 (1981) 425
\end{thebibliography}
\end{document}